\documentstyle{article}
 \def\be{\begin{equation}}
 \def\ee{\end{equation}}
 \def\bea{\begin{eqnarray}}
 \def\eea{\end{eqnarray}}
 \def\ba{\begin{array}}
 \def\ea{\end{array}}
 
 \def\nn{\nonumber}

\begin{document}


\title{{\bf New Vacuum State of the Electromagnetic Field-Matter Coupling System
and the Physical Interpretation of Casimir Effect}}

\author{LI Tong-Zhong (ÀîÍ­ÖÒ\hspace*{16mm})\footnote{ Ref. No. 23-614; Email:
tzli@sjtu.edu.cn; Tel: 021-54744134(H); 021-54747927(O);
021-54742908(O); Fax: 021-54741040} }
\maketitle {\small {\it
Department of Physics, Shanghai Jiao Tong University, Shanghai
200240, P. R. China}}


 A new concise method is presented for the calculation of the
ground-state energy of the electromagnetic field and matter field
interacting system. With the assumption of squeezed-like state, a
new vacuum state is obtained for the interacting system. The
energy of the new vacuum state is lower than that given by the
second-order perturbation theory in existing theories. In our
theory, the Casimir effect is attributed neither to the quantum
fluctuation in the zero-point energy of the genuine
electromagnetic field nor to that in the zero-point energy of the
genuine matter field, but to that in the vacuum state of the
interacting system. Both electromagnetic field and matter field
are responsible for the Casimir effect.




    The phenomenon that an attractive force exists between two
uncharged conducting metal plates is called the Casimir effect
because of the celebrated formula that Casimir derived for the
force between the two plates.$^{\cite{Casimir}}$ Casimir
calculated the sum of the zero-point energies of the normal modes
of the electromagnetic (EM) fields and showed that the sum and the
Casimir force, as the spatial derivative of the total energy,
depend on the distance between the two plates.$^{\cite{Casimir}}$
Later, Lifshitz$^{\cite{Lifshitz}}$ and van Kampen \textit{et
al.}$^{\cite{Kampen}}$ made the generalization of Casimir effect
to the case of two dielectric plates.

    The origin of the Casimir effect has usually been attributed to
the quantum fluctuations in the EM zero-point energy due to the
presence of the two plates. However, Schwinger and his
cooperators$^{\cite{Schwinger1,Schwinger2}}$ proposed the
Schwinger's source theory of the Casimir effect and showed that
the Casimir force can be derived with no explicit reference to the
zero-point energy fluctuations of the EM field. Milonno and
Shih$^{\cite{Milonni}}$ developed a new source theory within the
framework of completely conventional quantum electrodynamics. In
this theory, the Casimir force can be derived in terms of the
quantum fluctuations of atomic dipoles in the dielectric where the
EM field mediates the EM interactions between those dipoles.
Recently, Koashi and Ueda$^{\cite{Koashi}}$ published the
matter-field theory of Casimir force in which the matter and EM
field can be treated on an equal footing. Their strategy is to
explicitly diagonalize the matter-field Hamiltonian, which is
quadratic in its dynamic variables. The result shows that both EM
field and matter field contributes to the zero-point energy and
therefore to the Casimir force.

       So it is unclear up to now what the physical interpretation of the
Casimir effect is. It seems like it is largely a matter of taste
whether we attribute the Casimir force to the quantum nature of
the EM field or to that of the matter
field.$^{\cite{Milonni,Koashi}}$ However, according to the
postulates of quantum mechanics, to any well-defined observable
(or dynamical variable) in physics, there corresponds an operator
such that measurement of the observable yields values which are
eigenvalues of the corresponding operator. Therefore, any physical
effect should have unambiguous interpretations. On the other hand,
the interpretation of the physical effect should reflect the
physical reality.

      In this letter, we will treat the EM field and the matter
collective modes (plasmons in dielectric plates) on the same basis
as in Ref. \cite{Koashi}. A new concise method will be presented
for the calculation of the ground-state energy of the EM
field-matter interacting system. A qualitative interpretation of
the Casimir effect will be given based on the new vacuum state
obtained in this letter. In our theory, the Casimir effect is
attributed neither to the quantum fluctuations in zero-point
energy of the genuine EM field nor to that in zero-point energy of
the genuine matter field, but to that in zero-point energy of the
combined vacuum state of the EM field and the matter field.

    We now consider the interacting system of EM field and the matter
field. In quantum mechanics, the EM field can be described as a
set of harmonic oscillators corresponding to the normal mode
frequencies, which are determined from the Maxwell equations with
proper boundary conditions. The so-called matter field represents
the collective motions of various kinds of charged particles in
the dielectric substance.$^{\cite{Koashi}}$  They are surface mode
plasmons that can also be described as a set of harmonic
oscillators. Considering the interactions between these two sets
of harmonic oscillators, the Hamiltonian of the system is as
follows:
\begin{equation}
H = H_{0} + H_{I},
\end{equation}
where
\begin{equation}
 H_{0} \!= \!\!\sum\limits_{k}^{} {\hbar \omega _{a,k} \left( {a_{k} ^{ +
}a_{k} \!+ \frac{{1}}{{2}}} \right)} + \sum\limits_{i}^{} {\hbar
\omega _{b,i} \left( {b_{i} ^{ +} b_{i} + \frac{{1}}{{2}}}
\right)}
\end{equation}
and $H_{I} $ is the interaction energy of linearly interacting
subsystems, $\hbar $ the Planck constant. $a_{k} ^{ +} $($a_{k} $)
represents the creation (annihilation) operator of EM field with
the wavevector \textit{k} and frequency $\omega _{a,k} $, and
$b_{i} ^{ +} $ ($b_{i} $) the creation (annihilation) operator of
the collective modes in the dielectric substance characterized by
frequency $\omega _{b,i} $. $a_{k}$, $a_{k} ^{ +}$, $b_{i}$ and
$b_{i} ^{ +} $ are Bosonic operators and satisfy the relations $
\left[ {a_{k} ,a_{k'} ^{ +} } \right] = \delta _{k,k'},
\left[{b_{i},b_{j} ^{ +} } \right] = \delta _{i,j}$ and $
 \left[ {a_{k} ,b_{j}}  \right] = \left[ {a_{k} ,b_{j} ^{ +} } \right] =
\left[ {a_{k} ^{ +} ,b_{j}}  \right] = \left[ {a_{k} ^{ +} ,b_{j}
^{ +} } \right] = 0,$ where $\delta _{i,j} $ and $\delta _{k,k'} $
are the Kronecker $\delta $-functions.

    In general, the ground-state energy of the system can be obtained by
directly diagonalizing the Hamiltonian.$^{\cite{Koashi}}$ However,
in order to obtain analytic results so as to have an explicit
interpretation of the Casimir effect, without losing generality,
we consider the interacting system with single mode of EM field
and single mode matter field. In this case, we have the following
simple form of Hamiltonian$^{\cite{Koashi}}$
\begin{equation}
H_{0} = \hbar \omega _{a} \left( {a^{ +} a + 1/2} \right) + \hbar \omega
_{b} \left( {b^{ +} b + 1/2} \right)
\end{equation}
and
\begin{equation}
H_{I} = v_{ +}  a^{ +} b^{ +}  + v_{ -}  a^{ +} b + v_{ +} ^{\ast}
ab + v_{ -} ^{\ast}  ab^{+}.
\end{equation}
where $a^{ +} $ ($a$) and $b^{ +} $ ($b$) are the creation
(annihilation) operators of the single mode EM field and matter
field respectively, while $v_{ +}$, $v_{ -}$, $v_{ +} ^{\ast}$ and
$v_{ -} ^{\ast}$ the coupling coefficients between the two
fields.$^{\cite{Koashi}}$

    From the Hamiltonian, Eq. (4), we know that the two subsystems are coupled
linearly with each other and the numbers of photons (of the EM
field) and plasmons (of the matter field) are not conservative
respectively. Therefore it is appropriate to write the
ground-state wave function for the interacting system in the form
of squeezed-like state as
follows:$^{\cite{Wangk1,Wangk2,Fan1,Fan2}}$
\begin{equation}
\left| {g} \right\rangle = A \exp\left( {\alpha a^{ +} a^{ +} +
\beta b^{ +} b^{ +}  + \gamma a^{ +} b^{ +} } \right)\left| {0,0}
\right\rangle ,
\end{equation}
where $\left| {0,0} \right\rangle $ is the unperturbed vacuum
state of the EM and matter fields with the property $a\left| {0,0}
\right\rangle = b\left| {0,0} \right\rangle = 0$ and $\alpha$,
$\beta$ and $\gamma $ are the parameters to be determined in the
following calculation. $A$ is a constant for the normalization of
the state  $\left| {g} \right\rangle $ according to the relation
$\langle g \left| {g} \right\rangle =1$. With the aid of closure
relations of coherent states of $a$ and $b$, we have \bea
A&=&\left \{
\!\!\int\!\!\frac{\textrm{d(Re}\phi)\textrm{d(Im}\phi)}{\pi}
\frac{\textrm{d(Re}\varphi)\textrm{d(Im}\varphi)}
{\pi}\!\exp\!\!\left[2\textrm{Re}(\alpha\phi\phi)\right.\right.
\nn\\
&& \left . \left.+\!2\textrm{Re}(\beta\varphi\varphi)
\!+\!2\textrm{Re}(\gamma\phi\varphi)\!-\!\phi^*\phi\!-\!\varphi^*\varphi\right]
\right\}^{-\frac{1}{2}} \nn \eea
 where $\phi$ and $\varphi$
are respectively eigenvalues of  $a$ and $b$, corresponding to
their coherent states $\left| {\phi} \right\rangle$ and $\left|
{\varphi} \right\rangle$, \textit{i.e.}, $a\left| {\phi}
\right\rangle =\phi\left| {\phi} \right\rangle $ and $b\left|
{\varphi} \right\rangle =\varphi\left| {\varphi} \right\rangle $.
 The state $\left| {g} \right\rangle $ satisfies the
following relations
\begin{equation}
  a\left| {g} \right\rangle = \left( {2\alpha a^{ +}  + \gamma b^{ +} }
\right)\left| {g} \right\rangle
\end{equation}
and
\begin{equation}
b\left| {g} \right\rangle = \left( {2\beta b^{ +}  + \gamma a^{ +}
} \right)\left| {g} \right\rangle .
\end{equation}
The assumed vacuum state $\left| {g} \right\rangle $ should be the
solution of the Schr\"{o}dinger wave equation and satisfy
\begin{equation}
  H\left| {g} \right\rangle = E\left| {g} \right\rangle ,
\end{equation}
from which the parameters $\alpha$, $\beta$ and $\gamma $ and
therefore the vacuum state $\left| {g} \right\rangle $ and the
corresponding energy $E$ can be determined.

    Putting Eq. (5) into Schr\"{o}dinger equation (8) and using Eqs.
(6) and (7), we get \bea
 0&\!\!=\!\!&+ v_{ + }^{\ast}  \gamma \left| {g} \right\rangle +
\frac{{1}}{{2}}\left( {\hbar \omega _{a} + \hbar \omega _{b}}
\right)\left| {g} \right\rangle -E\left| {g} \right\rangle \nn \\
 & &
+\!2\hbar \omega _{a} \alpha a^{ +} a^{ +} \left| {g}
\right\rangle + v_{ -} \gamma a^{ +} a^{ + }\left| {g}
\right\rangle +2v_{ + }^{\ast}  \alpha \gamma a^{ +} a^{ +} \left|
{g}  \right\rangle \nn \\
& &+ 2\hbar \omega _{b} \beta b^{ + }b^{ +} \left| {g}
\right\rangle +v_{ -} ^{\ast} \gamma b^{ +} b^{ +} \left| {g}
\right\rangle +2v_{ +}^{\ast} \beta\gamma  b^{ +} b^{ +} \left|
{g} \right\rangle \nn
\\
 &&+\! \left(\hbar \omega _{a}+\! \hbar \omega _{b}\right)\gamma a^{
+} b^{ +} \left| {g} \right\rangle \!\!+\!\! v_{ +} a^{ +} b^{ +}
\left| {g}  \right\rangle\!\!+\!\!v_{ +} ^{\ast}  \gamma ^{2}a^{
+} b^{ +} \left| {g}
\right\rangle    \nn \\
 && + 2v_{ -} ^{\ast}
\alpha a^{ +} b^{ +} \left| {g} \right\rangle\!+\!2v_{ -} \beta
a^{ +} b^{ +} \left| {g} \right\rangle \!+\! 4v_{ +} ^{\ast}
\alpha \beta a^{ +} b^{ +} \left| {g} \right\rangle . \eea
\noindent In the above equation, we have put those terms, such as,
$\left| {g} \right\rangle$, $a^{ + }a^{ +} \left| {g}
\right\rangle$, $b^{ +} b^{ +} \left| {g} \right\rangle$ and $a^{
+} b^{ +} \left| {g} \right\rangle $ together. In order to get a
solution of the Schr\"{o}dinger equation, we may let the
coefficients of those terms equal zero
respectively.$^{\cite{Wangk1,Wangk2,Li}}$  Then we have
\begin{equation}
E = v_{ +} ^{\ast}  \gamma + \frac{{1}}{{2}}\left( {\hbar \omega
_{a} + \hbar \omega _{b}}  \right),
\end{equation}
\begin{equation}
2\hbar \omega _{a} \alpha + v_{ -} \gamma + 2v_{ +} ^{\ast} \gamma
\alpha = 0,
\end{equation}
\begin{equation}
2\hbar \omega _{b} \beta + v_{ -} ^{\ast}  \gamma + 2v_{ +}
^{\ast}  \gamma \beta = 0,
\end{equation}
\begin{equation}
\left( {\hbar \omega _{a}\! +\! \hbar \omega _{b}} \right)\gamma
\!+\! v_{ +} \! +\! v_{ +} ^{\ast}  \gamma ^{2} \!+ \!2v_{ -}
^{\ast} \alpha\!\!+\!\! 2v_{ -} \beta \!+\! 4v_{ +} ^{\ast} \alpha
\beta \!\!=\!\! 0.
\end{equation}
\noindent In this case we can get a solution with above conditions
although we could not assert that it is a unique one. We know that
$\left| {g} \right\rangle$, $a^{ + }a^{ +} \left| {g}
\right\rangle$, $b^{ +} b^{ +} \left| {g} \right\rangle$  and $a^{
+} b^{ +} \left| {g} \right\rangle $ are not orthogonal to each
other but they are linearly
 independent. All the states with  $(a^+)^m(b^+)^n\left| {g} \right\rangle$  build
a complete base  of Hilbert space. Therefore the solution of Eqs.
(10)-(13) should be a state of the system. In addition, we will
see below that the
 state  energy  is lower than that of the ordinary vacuum
state. This fact means that the new state could  be regarded as
the physical vacuum state of the given system.

     From Eqs. (10)-(12) we can get the relations among the parameters
$\alpha$, $\beta$ and $\gamma $ and the state energy $E$.
Considering the following confining conditions derived from the
normalization of the state $\left| {g} \right\rangle $
\begin{equation}
\left| {\alpha} \right| < 1/2, \quad \left| {\beta}  \right| <
1/2,\quad \left| {\gamma}  \right| < 1 ,
\end{equation}
we obtain the following unique possible energy value for the
ground state from equation (13): \bea E &\!\!=\!\!&
\frac{{1}}{{2}}\left \{ \left[ \left( \left ( \hbar \omega
_{a}\right)^{2} + \left( \hbar \omega _{b} \right)^{2} \right) -
2\left( \left| v_{ +}  \right|^{2} - \left| {v_{ -} } \right|^{2}
\right) \right] \right.
\nn \\
& & \!\!+\! 2 \left . \!\!\left[ \!\left( \!\hbar\omega _{a} \hbar
\omega _{b} \!\!-\!\! \left| v_{+} \right|^{2} \!\!+\!\! \left|
\!v_{-}\right|\!^{2} \right)\!^{2} \!\!- 4\left|
v_{-}\right|^{2}\left(\! \hbar \omega _{a} \hbar \omega _{b}
\right)\! \right]\!^{\frac{1}{2}}\right \}^{\frac{1}{2}} \eea
which is obviously lower than the unperturbed vacuum
state,\textit{ i.e}., $E < E_{0} = {{\left( {\hbar \omega _{a} +
\hbar \omega _{b}} \right)} \mathord{\left/ {\vphantom {{\left(
{\hbar \omega _{a} + \hbar \omega _{b}} \right)} {2}}} \right.
\kern-\nulldelimiterspace} {2}}$.

   For the weak interaction, \textit{i.e}., $\left| {v_{
+} } \right|$, $\left| {v_{ -} }  \right| < < \hbar \omega _{a}$,
and $\hbar \omega _{b}$ , we have \bea E &\!=\!&
\frac{{1}}{{2}}\left( {\hbar \omega _{a} + \hbar \omega _{b}}
\right) - \frac{{\left| {v_{ +} } \right|^{2}}}{{\left( {\hbar
\omega _{a} + \hbar \omega _{b}}
\right)}}\nn \\
& & -\frac{{\left| {v_{ +} }  \right|^{2}}}{{2\left( {\hbar \omega
_{a} + \hbar \omega _{b}} \right)}}\left[ {\frac{{2\left| {v_{ +}
} \right|^{2}}}{{\left( {\hbar \omega _{a} + \hbar \omega _{b}}
\right)^{2}}} + \frac{{\left| {v_{ -} } \right|^{2}}}{{\left(
{\hbar \omega _{a} \hbar \omega _{b}} \right)}}} \right]
 \eea
to the fourth order of ${\raise0.7ex\hbox{${\left| {v_{ \pm} }
\right|}$} \!\mathord{\left/ {\vphantom {{\left| {v_{ \pm}
}\right|}{\hbar\omega_{a,b}}}}\right.\kern-\nulldelimiterspace}\!\lower0.7ex\hbox{${\hbar
\omega _{a,b} }$}}$.  This result agrees exactly with that
obtained by Koashi and Ueda$^{\cite{Koashi}}$ to the second order
of ${\raise0.7ex\hbox{${\left| {v_{ \pm} }  \right|}$}
\!\mathord{\left/ {\vphantom {{\left| {v_{ \pm} }  \right|} {\hbar
\omega _{a,b}}
}}\right.\kern-\nulldelimiterspace}\!\lower0.7ex\hbox{${\hbar
\omega _{a,b}} $}}$ and it is lower than that in Ref.\cite{Koashi}
to the fourth order. At the same time, our theory applies not only
for the weak interacting system but also for the relatively strong
interacting system. We can get the analytic result for strong
interacting system.

   We now discuss the physical interpretation of the Casimir effect.
Up to now, there are two quite different kinds of physical
interpretations of the Casimir force,\textit{ i.e}., the
field-theory$^{\cite{Casimir,Lifshitz}}$ and the source
theory.$^{\cite{Kampen,Schwinger1,Schwinger2}}$ In the
field-theory, the Casimir force is attributable to the zero-point
energy fluctuations of the genuine EM field, where the existence
of the conducting plates can only change the boundary conditions
of the EM field. However, Schwinger \textit{et
al}$^{\cite{Schwinger1,Schwinger2}}$ got the same formula of the
Casimir force based on the source theory, and they pointed out
that the Casimir effect is attributable to the quantum
fluctuations of the genuine matter field, where the EM field is
only the medium to transport the electromagnetic interactions
between atomic dipoles in the dielectric
substances.$^{\cite{Milonni}}$ However, if we choose appropriate
ordering of operators in the Hamiltonian of linearly coupled
harmonic oscillators, the second-order correction to the
ground-state energy can therefore be attributed solely to quantum
fluctuations of either one of the oscillators, the EM field
operator or the matter field one.$^{\cite{Milonni,Koashi}}$ Then
the Casimir force can be attributable either to the genuine EM
field or to the genuine matter field.$^{\cite{Koashi}}$ According
to the theories of Casimir,$^{\cite{Casimir}}$ Schwinger
\textit{et al.}$^{\cite{Schwinger1,Schwinger2}}$ and Koashi and
Ueda,$^{\cite{Koashi}}$ it seems as if the physical interpretation
of the Casimir force depends on what physical models or
mathematical methods we used, but not on the expression of
physical reality.

    In this letter, we used a new concise method to calculate the ground-state
energy of the EM field and matter interacting system. The obtained
vacuum energy is obviously lower than that of the unperturbed
vacuum state and that given by the second-order perturbation
theory.$^{\cite{Koashi}}$ So the new state can be regarded as the
real vacuum state of the interacting system. We may use this state
to interpret the source of the Casimir force. In the discussion,
we have dealt with the EM field and the matter field on the same
basis. This is very important to the interpretation of the Casimir
force because relevant physical observables are expressed in terms
of eigenvalues and eigenstates of the full Hamiltonian of the
entire system$^{\cite{Koashi}}$ according to postulates in quantum
mechanics.

        In the Hamiltonian of the interacting system, i.e.,
$H_{0} + H_{I} $ (Eqs. (3) and (4)), there are two points in
connection with the distance ($d$) between the two dielectric
plates. On the one hand, the surface plasmon (such as its
frequency $\omega _{b} $) does depend both on the parameters of
the dielectric surface and on that of the space between the two
plates, and therefore it must be related to the distance $d$. On
the other hand, the EM mode (such as the frequency $\omega _{a} $)
is affected by the boundary condition and dielectric constant and
therefore is related to the distance $d$ too. The EM field can
alter the properties of surface plasmon (such as the frequency
$\omega _{b} $) by the interaction between the two fields. For
example, this interaction exists only in a thin film beneath the
surface of a perfectly conducting metal. Therefore we know that
all the physical quantities, such as $\omega _{a} $, and $ \omega
_{b} $, and all the coupling coefficients $v_{ +}$, $v_{ -}$, $v_{
+} ^{\ast}$ and $v_{ -} ^{\ast}$ in the Hamiltonian (3) and (4)
depend on the distance ($d$) between the two dielectric plates. In
summary, the vacuum state (Eq. (5)) and the eigenenergy $E$ (Eq.
(15)) of the interacting system depend on the distance \textit{d}.

      Because the Casimir force is derived from the change in the
ground-state energy (\textit{E}) with respect to a virtual
infinitesimal displacement of the dielectric plates,
\textit{i.e}., $f_{\rm Casimir} = - \frac{{\partial E}}{{\partial
d}}$, we have the conclusion that Casimir effect is attributable
to the quantum fluctuations of the combined vacuum state of the EM
field and the matter field. It is the interaction between the two
fields that produces the Casimir effect.

     In summary, we have presented a new concise method for the calculation
of the ground-state energy of the interacting system of the EM
field and matter field. With the assumption of squeezed-like
state, a new vacuum state is obtained for the interacting system.
The energy of the vacuum state is lower than that given by the
second-order perturbation theory.$^{\cite{Koashi}}$ Based on the
new vacuum state, the Casimir effect is attributed to the
fluctuations in zero-point energy of the vacuum state of the
interacting system. Both EM field and matter field are responsible
for the Casimir effect.

\begin{description}

\item{Casimir} Casimir H B G 1948 {\it Proc. K. Ned Akad. Wet.} B
{\bf 51} 793
\item{Lifshitz} Lifshitz E M 1956 {\it Sov. Phys.
JETP} {\bf 2} 73
 \item{Kampen}  van Kampen N G, Nijboer B R A and
Chram K 1968 {\it Phys. Lett.} A {\bf 26} 307 \item{Schwinger1}
Schwinger J, DeRaad L L and Milton K A 1978 {it Ann. Phys.} {\bf
115} 1 \item{Schwinger2} Schwinger J 1975 {\it Lett. Math. Phys.}
{\bf 1} 43 \item{Milonni} Milonni P W and Shih M L 1992 {\it Phys.
Rev.} A {\bf 45} 4241 \item{Koashi} Koashi M and Ueda M 1998 {\it
Phys. Rev.}  A {\bf 58} 2699 \item{Wangk1} Wan S L and Wang K L
2000 {\it Chin. Phys. Lett.} {\bf 17} 129 \item{Wangk2}Wan S L and
Wang K L 2003 {\it Chin. Phys. Lett.} {\bf 20} 117 \item{Fan1}Fan
H Y and Liu N L 2001 {\it Chin. Phys. Lett.} {\bf 18} 322
\item{Fan2}Fan H Y and Fan Y 2002 {\it Chin. Phys. Lett.} {\bf 19}
159 \item{Li} Li T Z, Wang K L, Qin G and Yang J L 1998 {\it J
Phys.}: CM {\bf 10} 319

\end{description}

\end{document}